\documentclass[baaa]{baaa}

\usepackage[pdftex]{hyperref}
\usepackage{subfigure}
\usepackage{natbib}
\usepackage{helvet,soul}
\usepackage[font=small]{caption}

\begin{document}


\journalvol{58}
\journalyear{2015}
\journaleditors{P. Benaglia, D.D. Carpintero, R. Gamen \& M. Lares}


\contriblanguage{1}


\contribtype{1}

\thematicarea{9}

\title{FOTOMCAp: a new quasi-automatic code for high-precision photometry}


\titlerunning{FOTOMCAp: a new quasi-automatic code for high-precision photometry}


\author{R. Petrucci\inst{1,2} \& E. Jofr\'e\inst{1,2}}
\authorrunning{Petrucci \& Jofr\'e}


\contact{romina@oac.unc.edu.ar}

\institute{Observatorio Astron\'omico de C\'ordoba (OAC) \and
  Consejo Nacional de Investigaciones Cient\'ificas y T\'ecnicas (CONICET)
}


\resumen{
La búsqueda de planetas parecidos a la Tierra utilizando la técnica de tránsitos ha impulsado el desarrollo de estrategias que permiten obtener curvas de luz cada vez más precisas. En este contexto desarrollamos el programa FOTOMCAp. Éste es un código cuasi-automático en lenguaje de programas IRAF que emplea el método de corrección por apertura y permite obtener curvas de luz de alta precisión. En esta contribución describimos cómo funciona el nuevo código y mostramos los resultados obtenidos para curvas de luz de tránsitos planetarios.}

\abstract{The search for Earth-like planets using the transit technique has encouraged the development of strategies to obtain light curves with increasingly precision. In this context we developed the FOTOMCAp program. This is an IRAF quasi-automatic code  which employs the aperture correction method and allows to obtain high-precision light curves. In this contribution we describe how this code works and show the results obtained for planetary transits light curves.}



\maketitle

\section{Introduction}
\label{S_intro}

\noindent
To date, more than 1900 extrasolar planets\footnote{\url{http://exoplanet.eu/catalog/}} have been discovered using different detection
methods. Almost 50 per cent of these new worlds were found by the Kepler satellite \citep{borucki}
through the techniques of transit \citep{charbo, henry} and Transit Timing Variations or TTV \citep{holman}.
The analysis of data provided by the mission has revealed a wide variety of planetary systems. Some of the most amazing
discoveries include the existence of Earth-like planets located in the habitable zone of their stars \citep{quintana, jenkins2015}, multi-planetary systems with more than one transiting planet \citep{masuda, becker}, 
or disintegrating minor planets orbiting a white dwarf \citep{croll}.

For ground-based observatories, the detection of light curves with shallow minima
caused by small transiting planets represents an enormous challenge. Besides the 
precautions taken to acquire the best possible data, it is highly desirable that the procedure to
obtain differential magnitudes ensures the determination of the most precise light curves. 
In this sense, almost all the groups carrying out systematic photometric follow-ups to find planets
have developed their own pipelines to analyse the photometric data. However, most of these codes are not publicly available.

In this contribution we present the quasi-automatic program FOTOMCAp, written in IRAF language, developed to determine precise instrumental magnitudes. In Section 2, we describe the code. In Section 3, we examine possible differences between the results obtained with FOTOMCAp and those with our previous code FOTOMCC. In Section 4 we compare several algorithms to determine the stellar centre and flux, and the sky level value. Finally, in Section 5 we present our conclusions.

\section{The FOTOMCAp code}
The aim of this program is to automatically measure instrumental magnitudes of several stars that belong to a same field for which a large number of images is available, which is the usual scenario in the study of planetary transits or variable stars. As input, this code requires: a) all the images of the field under study, b) the values of different parameters that characterize the employed CCD (i.e. readout noise, gain, and saturation level) and the telescope (i.e. primary mirror diameter and the elevation of the observatory above sea level), c) a file with information provided by the \texttt{daofind} task of an i-number of previously selected bright and isolated field stars, and d) a reference image. This reference image is chosen by the user considering that it contains all the stars of interest and where each one presents an optimum level of ADUs. As last requirement, FOTOMCAp needs a shift estimation (in pixels) of the positions of the stellar objects with respect to those of the reference image.  Once the centres of all the stars in all the images are identified, the code carries out a sequence of fully automatic steps to measure the instrumental magnitude of each star in each image by applying the method of aperture correction \citep{howell, stetson}.
This method is performed by the program as follows:\\

1) Magnitudes of the selected bright and isolated stars are computed in two different ways: 
\begin{itemize}
\item{Growth curves: it consists in measuring different magnitudes for the same star by applying larger and larger apertures each time. The final magnitude adopted for the star, m${_\mathrm{CC}}$, will be the one that remains unchanged in spite of the fact that the aperture size continues to increase.}
\item{Signal-to-noise ratio (SN) curves: it consists in determining the value of SN (see Section 2.1) for the same star by applying larger and larger apertures each time. The final magnitude adopted for the star, m$_\mathrm{{SN}}$, will be that calculated considering the aperture size for which SN is maximum.}
\end{itemize} 

\noindent For the determination of both, m$_\mathrm{{CC}}$ and m$_\mathrm{{SN}}$, the code employs the \texttt{phot} task within the \texttt{DAOPHOT} package. The values of sky used to determine the stellar centre and full-width-half-maximum (FWHM) required by \texttt{phot}, are automatically measured by FOTOMCAp in the surrounding area of each star and considering the brightest stars in the field, respectively. The sky ring inner radius (annulus) is computed as the selected aperture size plus 5 pixels, its thickness (dannulus) is fixed in 5 pixels and the sky level value inside it is calculated by the algorithm MODE (see Section 4).\\  

2) For each one of these i-stars, the program computes the $\mathrm{\Delta}$m$_\mathrm{{i}}$=(m$_\mathrm{{CC}}$ $-$m$_\mathrm{{SN}}$)$_\mathrm{{i}}$ difference, and for every image obtains a $\mathrm{\Delta}$m value calculated as the median of the $\mathrm{\Delta}$m$_\mathrm{{i}}$. The $\mathrm{\Delta}$m value determined for a given image constitutes the aperture correction for that image.\\

3) Finally, the instrumental magnitude, m$_\mathrm{{ins}}$, of every star in the field is obtained as m$_\mathrm{{ins}}$=m$_\mathrm{{SN}}$+$\mathrm{\Delta}$m for each image, where m$_\mathrm{{SN}}$ is computed as we mentioned in 1). The adopted errors are those computed by the \texttt{phot} task.\\

\noindent Once FOTOMCAp stops running, it creates three files: one containing relevant information, including the instrumental magnitudes and errors of all the stars in each image, and the other two with the sky level and FWHM values measured for all the images.

\subsection{Determination of SN}

The code automatically computes the SN through the expression given by \cite{merline},

\begin{equation*}
SN=\frac{N_{\star}}{\sqrt{N_{\star}+n_{\text{pix}}\Bigg(1+\frac{n_{\text{pix}}}{n_{\text{B}}}\Bigg)(N_{\text{S}}+N_{\text{D}}+N^{2}_{\text{R}})+\sigma^{2}_{\text{S}}}},
\end{equation*}

\noindent where $N_{\star}$ is the total number of photons collected from the star inside the \textbf{$n_\mathrm{pix}$} pixels that constitute the used aperture. $N_{S}$, $N_{D}$ and $N_{R}$ are the photons per pixel coming from the sky, the dark current, and the readout noise, respectively. In this equation it is assumed that the sky has been estimated over an annulus covering $n_{B}$ pixels. The $\sigma_{S}$ term represents the contribution from the atmospheric scintillation given by \cite{dravins},\\

\begin{equation*}
\sigma_{S}=0.09N_{\star}\frac{X^{3/2}}{D^{2/3}\sqrt{2t_{\text{exp}}}} e^{(-h/8)},
\end{equation*}

\noindent where $X$ is the airmass, $h$ the observatory altitude in km, $D$ the diameter of the telescope in cm, and \textbf{$t_\mathrm{exp}$} is the exposure time in seconds.

\section{Comparison with the FOTOMCC code}

FOTOMCC \citep{petrucci2013, petrucci2015} is a program written in IRAF language, developed to obtain high-precision light curves. Unlike FOTOMCAp, FOTOMCC determines instrumental magnitudes only employing the technique of growth curves. In Figure 1, we show the light curve of the same planetary transit obtained with the FOTOMCAp (circles) and FOTOMCC (squares) codes. It is important to mention that the differential magnitudes were determined, in both cases, following the procedure described in \cite{petrucci2013}. 

In Table 1, we present the standard deviation ($\mathrm{\sigma}$), useful number of images (N$_\mathrm{{ima}}$), and computer time employed to carry out the photometry (T$_\mathrm{{comp}}$) of the planetary transit obtained with the two codes.
The comparison between these values indicates that, in several important aspects, FOTOMCAp considerably improves the results that are obtained with FOTOMCC. For example we can mention an increase in the number of useful images to construct the light curve and a significant decrease in the standard deviation. The only drawback is that the FOTOMCAp code requires up to four times more computing time than FOTOMCC.

\begin{figure}[!ht]
  \centering
  \includegraphics[width=0.5\textwidth]{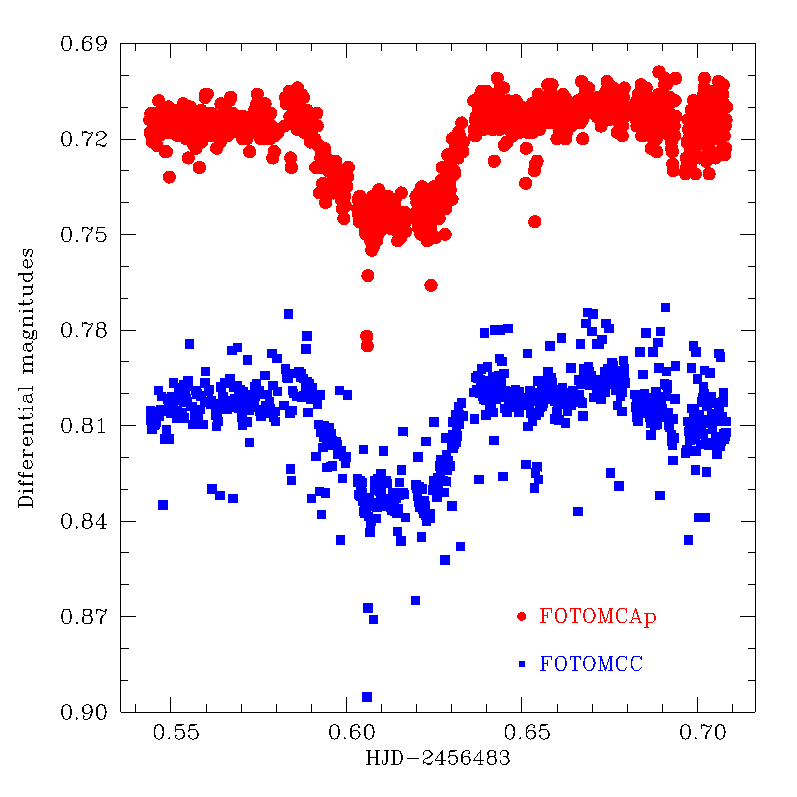}
  \caption{Light curves of the same transit event obtained with FOTOMCAp (circles) and FOTOMCC (squares). The light curves were shifted in the Y axis for better visualization.}
  \label{F_toobig}
\end{figure}

\begin{table}[!ht]
\centering
\caption{Comparison between the results obtained with the FOTOMCAp and FOTOMCC codes.}
\begin{tabular}{lccc}
\hline\hline\noalign{\smallskip}
\!\!	Code	&	$\mathrm{\sigma}$ (mag)	&	N$_\mathrm{{ima}}$	&	T$_\mathrm{{comp}}$ (hs)	\\
\hline\noalign{\smallskip}
\!\!	FOTOMCAp	&	0.013	&	901	&	12	\\
\!\!	FOTOMCC	&	0.021	&	852	&	3	\\
\hline
\end{tabular}
\end{table}

\section{Comparison between the algorithms to determine the stellar centre and flux, and the sky level value}

The \texttt{phot} task, used to measure instrumental magnitudes, has different algorithms to accurately determine the stellar centre location, to compute the star's flux, and to calculate the background sky value. 
To establish the most precise algorithm, we compared in each case the standard deviations ($\mathrm{\sigma}$) of the resulting light curves for a bright star and for a faint one (Table 2). Both objects belong to the same field that the star presenting the planetary transit shown in Figure 1.

As it can be seen, the differences between the values of $\mathrm{\sigma}$ are at most $\sim$ 0.2 mmag for the brightest stars and as much as $\sim$ 7 mmag for the faintest ones. However, it is important to notice that for the faint stellar objects these differences appear when the sky value is fitted by the CONSTANT algorithm. If we exclude the results obtained with this algorithm, the differences between the values of $\mathrm{\sigma}$ are at most $\sim$ 3 mmag. Taking into account that what the CONSTANT option does is to fix the sky to a value given by the user, which in this case is the mean of the values of sky automatically measured by FOTOMCAp, this result points out the need of using more complex methods to fit the background level when precise light curves are expected for faint stars. 

These results would indicate that, except for the CONSTANT algorithm to determine the sky value, the rest of the algorithms used to center the star, to calculate the stellar flux, and to measure the sky background level would not have a significant influence on the computed instrumental magnitudes.


\begin{table}[!ht]
\centering
\caption{Standard deviations in magnitude units of the light curves for a bright and a faint star, considering different algorithms to determine the centring, stellar flux and sky level.}
\begin{tabular}{lcc}
\hline\hline\noalign{\smallskip}
\!\!Algorithm & \!\!\!\! $\sigma_\mathrm{{bright}}$ & \!\!\!\!$\sigma_\mathrm{{faint}}$\\
\hline
\multicolumn{3}{c}{CENTRING} \\	
\hline
\!\!	CENTROID	&	0.0072	&	0.035	\\
\!\!	GAUSS	&	0.0071	&	0.035	\\
\!\!	NONE	&	0.0071	&	0.035	\\
\!\!	OFILTER	&	0.0071	&	0.035	\\
\hline		
\multicolumn{3}{c}{STELLAR FLUX} \\	
\hline
\!\!	CONSTANT	&	0.0072	&	0.036	\\
\!\!	CONE	&	0.0072	&	0.035	\\
\!\!	GAUSS	&	0.0072	&	0.036	\\
\hline
\multicolumn{3}{c}{SKY LEVEL} \\	
\hline
\!\!	CENTROID	&	0.0072	&	0.037	\\
\!\!	CONSTANT	&	0.0073	&	0.042	\\
\!\!	CROSSCOR	&	0.0072	&	0.036	\\
\!\!	GAUSS	&	0.0071	&	0.038	\\
\!\!	MEAN	&	0.0072	&	0.035	\\
\!\!	MEDIAN	&	0.0072	&	0.036	\\
\!\!	MODE	&	0.0072	&	0.036	\\
\!\!	OFILTER	&	0.0072	&	0.036	\\
\hline
\end{tabular}
\end{table}

\section{Conclusions}
In this contribution we present a new quasi-automatic code, FOTOMCAp, developed in IRAF language. The purpose of this code is to measure instrumental magnitudes of several stars that belong to the same field, by using the method of aperture correction. 
Our results indicate that FOTOMCAp not only allows to compute light curves with standard deviations smaller than the ones obtained with the FOTOMCC program, but also permits to employ a larger number of images than the one used in our previous code.
The only noticeable drawback is that the computation time used by FOTOMCAp is $\sim$ 4 times larger than the one used by FOTOMCC. However, about this last point, we are currently working to improve the code efficiency and, in that way, significantly reduce the internal calculation time. 

\begin{acknowledgement}
R. P. and E. J. acknowledge the financial support from CONICET in the form of postdoctoral fellowships. We also thank the financial support provided by the local organization committee to attend the 58th Meeting of the Argentine Astronomical Society.
Finally, we thank the anonymous referee for his/her useful comments and suggestions, which helped to improve the quality of this paper.

\end{acknowledgement}


\bibliographystyle{baaa}
\small
\bibliography{biblio}
 
\end{document}